\documentclass[runningheads]{svmult}

\usepackage{makeidx}   
\usepackage{graphicx}  
\usepackage{subeqnar}  
\usepackage{multicol}  
\usepackage{physprbb}  
\makeindex             

\begin{document}
\title*{Globular Cluster Systems of Spirals}
\toctitle{Globular Cluster Systems of Spirals}
\titlerunning{GCSs of Spirals}
\author{Pauline Barmby}
\authorrunning{Barmby}
%
%
\institute{Harvard--Smithsonian Center for Astrophysics\\
60 Garden St., MS65\\
Cambridge, MA 02138 USA}

\maketitle              

\begin{abstract}

Recent observational results on the globular cluster systems
of spiral galaxies are summarized. Although the number of spirals
with well--studied GCSs is still small, new studies promise 
to increase it rapidly in the next few years. New telescopes 
and technology have contributed to increasingly detailed 
studies of GCs in Local Group galaxies like M31 and M33,
and more distant spirals like M81 and M104 are finally getting
the attention they deserve. The Milky Way GCS
looks to be reasonably typical of spirals, and still has
a lot to tell us about the history of the galaxy.
\end{abstract}

\section{Introduction}

We live in a spiral galaxy, so the best--studied globular cluster system 
will likely always be the Milky Way's. To understand if the Milky Way 
GCS is typical, we need to study the globular cluster systems 
of other spirals. If spiral--spiral mergers are important in
elliptical galaxy and GCS formation, information about spiral
GCSs is needed to predict the properties of elliptical GCSs.
Spiral galaxies contain most of the star formation in the nearby
universe, so they can provide the links between the star formation we 
study nearby and the high--redshift kind which formed globular clusters
and galaxy halos. GCs can be observed at much greater distances than
individual extragalactic bulge or halo stars, so they remain one
of the best ways of studying the formation of late-type galaxies.

Despite the importance of spiral galaxy GCSs, they are poorly--studied
compared to those of ellipticals. This is unsurprising, since ellipticals
have more globular clusters (there are no \lq giant spirals' with thousands
of GCs like M87), and their clusters are easier to find since the galaxy
light can be subtracted much more easily. The largest populations of
nearby galaxies are in clusters like Virgo and Fornax,
where spiral galaxies are less common than in the field.
As I was preparing my talk, I searched the Astrophysics
Data System for the words \lq \lq spiral~globular'' in the titles of
refereed papers. The tenth result was titled  
\lq \lq Photographic Effective Wavelengths of Spiral Nebulae and Globular
Clusters'', and published in 1919 \cite{ll17}!

In this review, I will emphasize recent observational results on the
globular cluster systems of well--studied spiral galaxies. These are
(unfortunately) still few enough in number to be discussed individually.
M31 probably gets somewhat more emphasis than it should, mostly by
virtue of being my thesis galaxy! Some galaxies and topics are not
discussed here since they are covered elsewhere in these proceedings.
Theories of globular cluster system formation have mostly concentrated on
elliptical galaxies, so relatively little emphasis is placed on 
theoretical interpretations. At the end of the review, I summarize what
is known, and still unknown, about the globular cluster systems of spirals.

\section{Milky Way GCS}

The individual GCs in the Milky Way are still important objects of study,
and many reports of results can be found in the proceedings of a meeting
(\lq \lq New Horizons in Globular Cluster Astronomy'', Padua) held shortly
before the ESO workshop. An excellent review of the Milky Way GCS can be found in
W. Harris' Saas--Fee lectures \cite{h00}. Here I mention some highlights
of new results on  the Milky Way globular cluster system.

One important question is: what is the total population of the Milky Way GCS?
The June 1999 edition of the Harris catalog \cite{h96} lists 147 objects.
Several recent estimates place the total number of GCs at about 160
\cite{h00,bbo98}. Two GCs hidden in the Galactic plane were recently 
discovered in the 2MASS database \cite{hur00}, and
other searches for star clusters in the Milky Way are underway \cite{db01}.

\begin{figure}[h]
\begin{center}
\includegraphics[width=.8\textwidth]{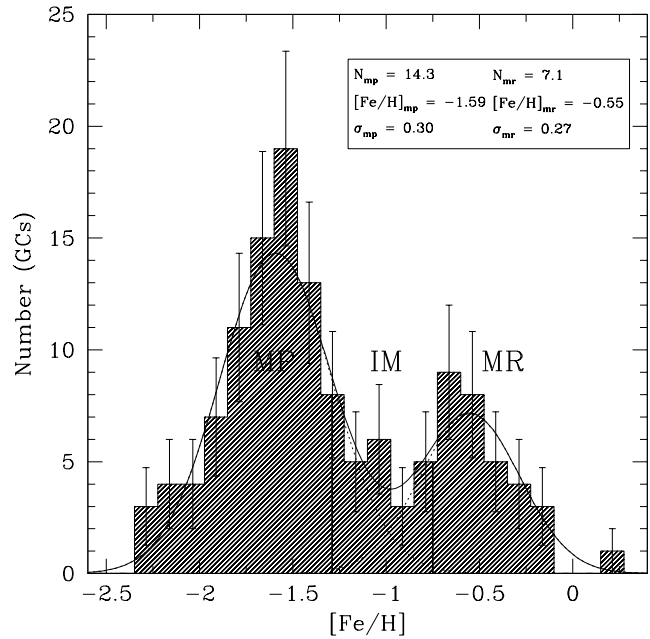}
\end{center}
\caption[]{Metallicity distribution of Milky Way globular clusters, from \cite{cot99}}
\label{cotefig}
\end{figure}

The issue of bimodality in GCS color distributions was discussed
extensively at this workshop. The Milky Way is one of the few galaxies
where a clear bimodality in metallicity (as opposed to color) can
be seen; a nice example of this is Fig.~\ref{cotefig}, from a paper by C\^ot\'e
\cite{cot99}.
He also shows that there is a clear difference in the kinematics
of the two metallicity groups, and suggests that the metal--rich clusters
are more likely to be associated with the Galactic bulge or bar,
rather than the disk.

\begin{figure}[h]
\begin{center}
\includegraphics[width=.8\textwidth]{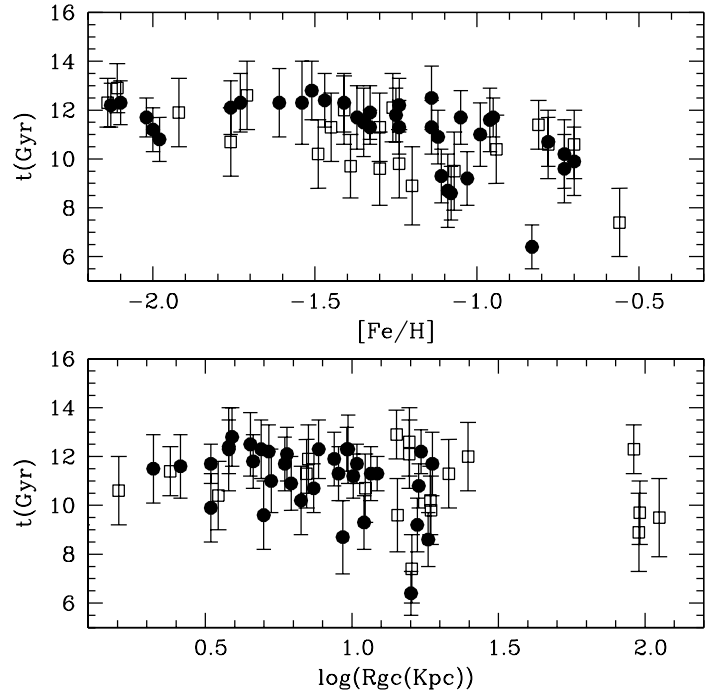}
\end{center}
\caption[]{Age distribution of Milky Way globular clusters, from \cite{sw02}}
\label{swfig}
\end{figure}

Another recent development is the compilation of several large databases
of color--magnitude diagrams for Milky Way GCs, including 
a database of HST $BV$ CMDs for nearly half the MW GCS \cite{pio02}, and
ground--based CMDs for large numbers of clusters \cite{ros99,sw02}.
The two groups with ground--based CMD databases have used them to 
investigate the age distribution of Milky Way globular clusters, with
similar results (see Fig.~\ref{swfig}).
The most metal--poor clusters ([Fe/H]$<-1.5$) are found to be coeval and old,
with an average age of about 12 Gyr. Intermediate--metallicity clusters 
($-1.5<{\rm [Fe/H]}<-1.0$) show a spread in ages of about 2 Gyr,
and the most metal--rich clusters are found to be coeval and somewhat younger than
the metal--poor ones. When age is compared to $R_{gc}$, the 
\lq outer halo' clusters are the ones with the age spread.
Some, but not all, of the age spread can be explained by
considering age as the \lq second parameter', that is, the
cause of different horizontal branch types in clusters of the same
metallicity.

Globular cluster destruction was discussed extensively at this meeting
(e.g., by Vesperini), and there is good observational evidence that
this process is actually taking place. A wide--field photographic imaging
study \cite{lmc00} finds that almost all of the 20 MW GCs studied show evidence
of interactions with the Galaxy (tidal tails, etc). A photometric
study using data from the Sloan Digital Sky Survey \cite{ode02} maps out 
the distribution of the cluster Palomar 5 on the sky -- nearly half its stars are
in the tidal tails. The large sky coverage and photometric stability of
the SDSS will no doubt enable many more such investigations.

\section{M31 GCS}

Although the M31 GCS has been extensively studied since Hubble \cite{hub32},
the size of the GCS is embarrassingly uncertain. Observational difficulties
are the primary reason: the M31 GCS covers a large area on the sky,
and the non--stellar appearance of M31 GCs makes them easily confused with
background galaxies. Several ongoing CCD surveys of M31 \cite{mas,lee01}
should be useful in clarifying the completeness of the photographic surveys. 
A model of the completeness of existing M31 GC surveys using HST data
yielded an estimate of the total number of M31 GCs as $460\pm70$ \cite{hst_1}.

The metallicity distribution of the M31 GCS shows clear evidence for 
bimodality \cite{b00,per02}, and Perrett (these proceedings) analyses
the difference in kinematics and spatial distribution between the two
metallicity groups. As with the Milky Way GCs, the age distribution of 
M31 GCs is much less well--known than the metallicity distribution.
By comparing the integrated colors of M31 GCs with stellar population
models, I concluded that the metal--rich GCs in M31 are younger than the metal--poor
clusters \cite{bh00}; but integrated photometry is a rather
blunt instrument, and spectroscopic and CMD studies are needed.
At least a few of the objects listed in the M31 GC catalogs are actually 
quite young \cite{wh01}.

The M31 globular cluster luminosity function (GCLF) has been studied 
for a long time, with the usual difficulties of correcting for incompleteness and
reddening. Many studies have shown that the luminosity function 
of the halo clusters is very similar to that in the Milky Way. 
With Huchra and Brodie, I attempted to compute luminosity functions for subsamples of the full
GCS, correcting for incompleteness and reddening \cite{bhb01}.
We found evidence for GCLF variation with both distance from the
galaxy center (due to destruction?) and metallicity (due to age?).
The GCLF variation is an effect not seen in other galaxies,
and to confirm it we intend to acquire the data necessary
to analyze the GCLF in the near--infrared, where reddening is less of a problem.

\begin{figure}[h]
\begin{center}
\includegraphics[width=.8\textwidth,angle=270]{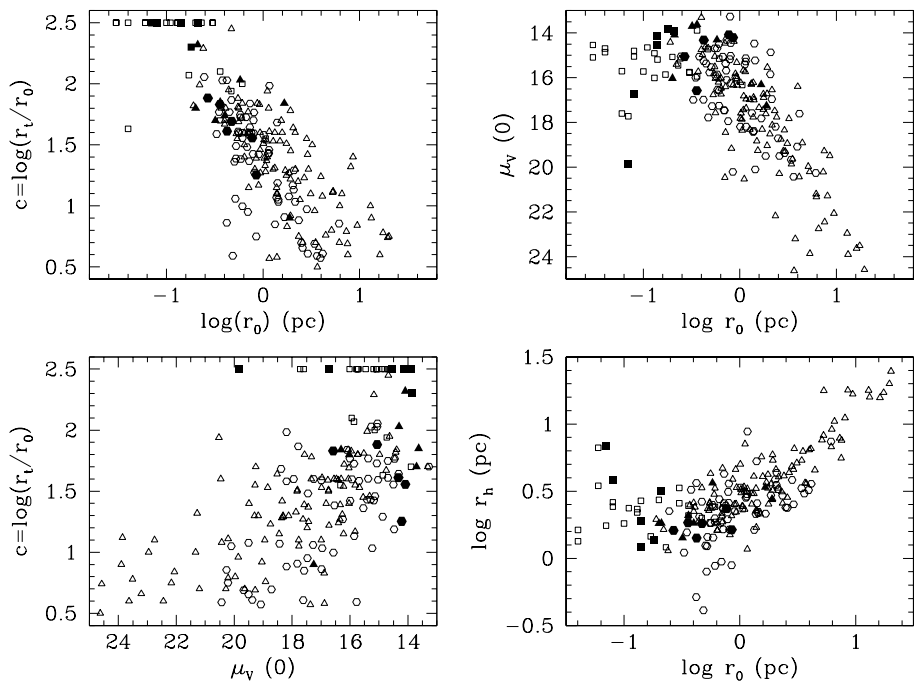}
\end{center}
\caption[]{Structural properties of M31 (hexagons) and Milky Way
(squares: core--collapsed, triangles: non-core--collapsed) GCs.
Clusters with X-ray sources are represented by filled symbols.
Adapted from data in \cite{hst_2} and \cite{xray01}}
\label{xrayfig}
\end{figure}

The Hubble Space Telescope has done a lot for the field of
extragalactic globular clusters. In M31 it allows both surface photometry
and color--magnitude studies, since the clusters are resolved into
individual stars. Measurements of structural parameters for $>70$ clusters from HST data
\cite{hst_2} showed that they have very similar properties to Milky Way clusters.
The two galaxies' clusters fall on similar trends in the \lq fundamental
plane' \cite{mcl00} of binding energy, luminosity, and concentration 
(see Fig.~\ref{xrayfig}), which implies
that they have very similar mass-to-light ratios, and probably similar
evolutionary histories. The M31 metal--rich clusters have slightly smaller sizes
than the metal--poor ones (see Sect.~\ref{m104sec}).

Other recent HST studies of M31 GCs include the work of
Clementini et al.\ \cite{clem01}, who reported the first detection of RR Lyrae 
stars in M31 GCs. Continuation of this work will provide
another \lq Population II' method of determining the distance to M31,
as well as allowing the study of the Oosterhoff phenomenon in another galaxy.
A detailed study of the large M31 cluster G1 \cite{mey01} shows it to be
similar to the Milky Way's $\omega$~Cen: it has a large ellipticity,
large mass and multiple stellar populations. Meylan et al.\
suggest that G1 is possibly the stripped nucleus of a dwarf elliptical 
galaxy; a similar idea has been put forward for $\omega$~Cen \cite{hr00}.
Zinnecker (this meeting) suggested that all GCs might be dwarf galaxy nuclei;
while this may not be a plausible way to form all GCs, perhaps it is
workable for the largest GCs in a galaxy.

The high spatial resolution of {\em Chandra} has allowed the definitive 
identification of X-ray sources in Local Group galaxies with their
optical counterparts. A survey of three ACIS fields in M31 \cite{xray01}
found that the most luminous X-ray sources are in GCs, and some of 
these are much more X-ray luminous than Milky Way GC X-ray sources.
The M31 clusters with X-ray sources tend to be optically brighter
(both in integrated magnitude and central surface brightness)
than those without; although not statistically significant,
there is also a hint that they are more concentrated (see Fig.~\ref{xrayfig}).
This agrees with the general idea that
the frequency of dynamical interactions determines the number
of X-ray binaries and that this scales with cluster central density.
The larger number of M31 X-ray GCs 
should allow a much more detailed comparison of cluster
and X-ray source properties.

\section{M33}

\begin{figure}[h]
\begin{center}
\includegraphics[width=.8\textwidth,angle=0]{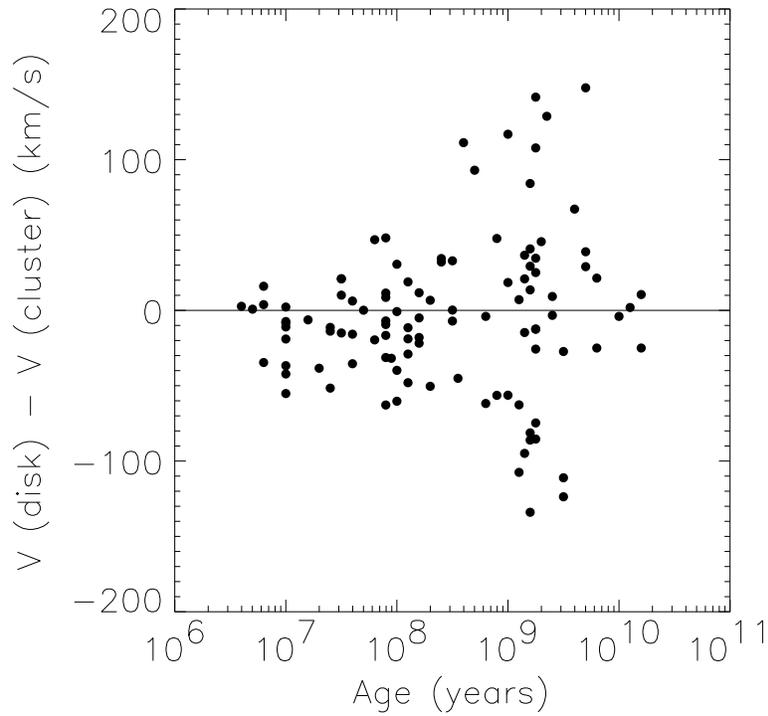}
\end{center}
\caption[]{M33 cluster velocities (relative to the disk) as a function
of age. From \cite{cbf5}}
\label{m33fig}
\end{figure}

Many spiral galaxies contain young, massive star clusters 
(Larsen, this meeting), and the Local Group spiral M33 is no 
exception. Chandar and collaborators have recently carried out an extensive 
study of the M33 cluster system \cite{cbf1}--\cite{cbf5}.
Using HST images, they identify many new star clusters in M33,
and estimate the total number of GCs at $75\pm14$. This
is a much larger value than previous estimates, and if correct
would give M33 a fairly large specific frequency (see Table~1).
It is clear that the the GCs have a much larger velocity dispersion than 
the young star clusters (see Fig.~\ref{m33fig}).
There are still relatively few M33 GCs with radial velocities, so
understanding their kinematics is difficult.  Modeling suggests that 
most belong to the halo rather than the disk \cite{cbf5}. 

The small number of M33 GC candidates known before the
Chandar et al.\ work have been heavily studied. HST imaging of 10 
halo clusters \cite{m33_cmd1} showed that 8 of them have red horizontal 
branches, an indicator of possible intermediate ($\sim7$~Gyr) ages. 
A spectrophotometric study \cite{ma3} yielded similar results. 
However, high resolution spectroscopy and HST surface photometry
of four of these clusters \cite{m33_hires} shows that they
have similar mass-to-light ratios to the Milky Way globulars
($\left<M/L_V\right>=1.5\pm0.2$), and fall on the same fundamental 
plane. This is puzzling since $M/L$ is expected to be lower
for younger clusters. Preliminary work on {\em Chandra} sources in
M33 \cite{mcd02} indicates that none are associated with GCs.
This might also argue against younger ages, since younger clusters
would have a higher turnoff mass and be more likely to have
luminous X-ray binaries (see \cite{xray01}).

\section{M104\label{m104sec}}

\begin{figure}[h]
\begin{center}
\includegraphics[width=.8\textwidth]{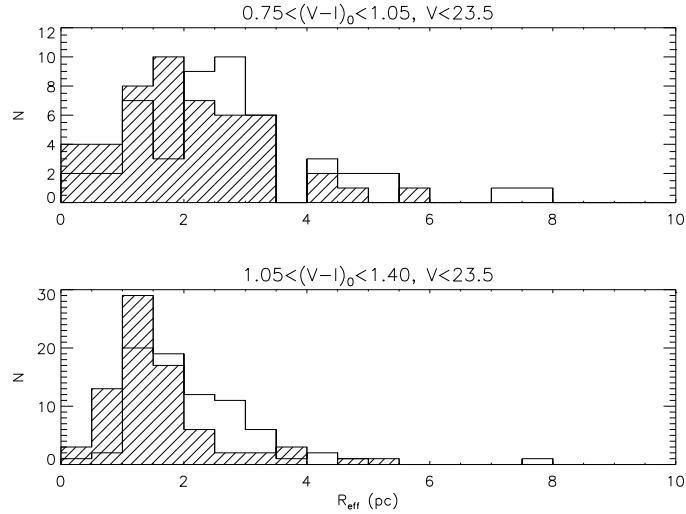}
\end{center}
\caption[]{Half--light radii for M104 clusters, from \cite{lar01b}.
The red clusters are smaller than the blue ones, an effect also seen
in other galaxies}
\label{m104fig}
\end{figure}

This meeting saw some debate about whether M104 (the \lq Sombrero') 
is a spiral galaxy with a large bulge, or an elliptical galaxy with a disk.
Since there are already many more elliptical galaxies than spirals
with GCS information, I consider it a spiral for this review.  
Recent works include HST imaging by Larsen et al.\ \cite{lar01b}: 
for three fields near the center of
M104, they find a total of $\sim170$ GC candidates and estimate
a total GCS population of about 1200$\pm600$.
The GC candidates have a bimodal color distribution: 45\% blue and 55\% red, 
somewhat different from the 65\%--35\% division seen in the Milky Way and M31.
There is a size difference between red and blue clusters (see Fig.~\ref{m104fig}),
although it is difficult to tell whether this is 
because the red clusters are located closer to the center of the galaxy.

Previous spectroscopy of the M104 clusters has so far included
fairly small samples of objects \cite{lar02,bri97}. Both groups
were able to estimate a mean metallicity for the cluster system
([Fe/H] $\sim -1.3$) and estimate the total mass of the galaxy.
Several groups presented new observations of M104 GCs at this meeting,
including VLT imaging (Moretti) and spectroscopy (Held),
and 2dF spectroscopy (Bridges). These new data should allow
the detailed kinematical studies not possible with previous
observations.

\section{M81}

Aside from a few papers in the mid-1990s \cite{pbh95,pr95}, the M81 GCS 
has received little attention until very recently. HST imaging of
8 fields \cite{cft1,cft2} revealed 114 new clusters, of which
about half were red enough to be old GCs. The total number of globular clusters 
was estimated at $\sim210$, with no bimodality detected in the GC color
distributions. Two groups have obtained spectra of M81 GC candidates
\cite{pbh95,sch02}; the latter group (Schroder et al.) studied the combined
sample of 44 GCs. The spectroscopic metallicities have large uncertainties,
so they did not formally test for a bimodal distribution.
Using colors as a rough metallicity indicator, Schroder et al. analyzed 
the kinematics of two groups, finding that the red clusters rotate with the disk, 
and the blue clusters show no rotation. The red clusters' rotation signature is
strongest for $4<R<8$~kpc, leading to the suggestion that these
clusters (and also their analogs in M31 and the Milky Way) comprise a \lq thick disk' 
population. This is at odds with other analyses of both the
Milky Way and M31 GCS kinematics, so perhaps a larger sample of
M81 GCs will tell a different story.

\section{NGC~4565, NGC~5907}

The clusters in these two edge-on galaxies were studied by
Kissler--Patig et al. \cite{kp99}, using $B$ and $I$ imaging from HST. Although 
these two galaxies have quite different structural properties,
their globular cluster systems appear to be quite similar.
Only small numbers of GCs were detected in each galaxy, so the
total number of clusters was estimated by comparing a model of the
selection effects to the spatial distribution of Milky Way clusters.
No bimodality was detected in the cluster candidate color distributions,
but the mean colors of clusters in NGC~4565 and NGC~5907 
were consistent with the mean metallicity of the MW GCS.
There is some indication that the NGC~5907 GCS is flattened,
of interest because this galaxy may also have a flattened halo.

\section{Other spirals}

A few less--well--studied spiral galaxies are collected here. Some
have new GCS studies in progress, while others were studied some
time ago and might benefit from a revisit with new telescope
and/or more sensitive detectors.

No detailed study of globular clusters has been done in 
the large nearby face--on spiral {\bf M101}, although a study of one HST/WFPC2
field \cite{bks96} detected 6 GC candidates and numerous young clusters.
An upcoming survey of M101 with the new ACS on HST (PI: K. Kuntz) should 
allow a much more thorough study. Another upcoming study is the thesis 
of K. Rhode, who has obtained extensive data on the GCSs
of several nearby edge-on spirals. These data will be extremely valuable for
comparison of the GCSs of spirals and ellipticals. In these proceedings,
Rhode discusses the GCS of {\bf NGC~7814}.

Also in these proceedings, Olsen discusses imaging of the GCSs
of spirals in the Sculptor group: {\bf NGC~55, 247, 253} and {\bf 300}. 
These galaxies are nearby, which makes their clusters bright, but
also means that GCs are more likely to be confused with background
galaxies. Beasley \& Sharples \cite{bs00} found that almost all of the NGC~253 and NGC~55 GC 
candidates they obtained spectra for were in fact galaxies.

Fischer et al.\ \cite{fis90} used early CCD images ($300\times500$ pixels)  
to study the GCS of {\bf NGC~5170}. They detected $130\pm20$ GC
candidates, and estimated the total population at $815\pm320$.
{\bf NGC~2403} is an M33--like galaxy in the M81 group; its globular
cluster system was studied by Battistini et al. \cite{b84}. They found evidence for 
a small population of old clusters and a somewhat larger population of
younger clusters (similar to M33).
{\bf NGC~2683}: a photographic study of this Sb spiral \cite{hhb85}
detected $100\pm31$ GCs, and estimated the total population at
$321\pm108$. 
Marginal detections of GCSs in the Virgo spirals {\bf NGC~4216} and
{\bf NGC~4569} \cite{han77} yielded population estimates of
$700\pm380$ and $1000\pm400$.

\section{Is the Milky Way GCS typical?}

\begin{table}
\caption{Globular Cluster Systems of Spiral Galaxies}
\begin{center}
\renewcommand{\arraystretch}{1.4}
\setlength\tabcolsep{5pt}
\begin{tabular}{lrcccll}
\hline\noalign{\smallskip}
galaxy& $M_V$ & $N_{GC}$ & $S_N$ & [Fe/H] & GCLF $V^0$ & Refs.\\
\noalign{\smallskip}
\hline
\noalign{\smallskip}
Milky Way& $-20.9$& $160\pm10$ & $0.7\pm0.1$ & $-1.6,-0.6$ & $-7.5$&\cite{h00}\\
M31      & $-21.4$& $460\pm70$& $1.2\pm0.2$ & $-1.4,-0.5$ & $-7.6$&\cite{hst_1,per02,bhb01}\\
M33      & $-18.9$&$75\pm14$ & $2.07\pm0.39$ & ??   & $-7.0$ & \cite{cbf4}\\
M81      & $-20.9$&$211\pm29$& $1.0\pm0.1$ &$-1.5\pm0.2$& $-7.5$ & \cite{pbh95,cft2}\\
M104     &$-22.1$&$1200\pm600$&$1.6\pm0.8$ & $-1.4,-0.5$ & $-7.6$ &\cite{lar01b}\\ 
NGC 4565 &$-21.4$&$204\pm38$ & $0.56\pm0.15$& $-1.3$& & \cite{kp99}\\
NGC 5907 &$-21.2$&$170\pm41$ & $0.56\pm0.17$& $-1.3$&  & \cite{kp99}\\
NGC 7814 &$-21.3$&230&0.7& & & Rhode\\
NGC 55   &$-19.1$&$8+ $&$>0.18$& & & Olsen\\
NGC 253  &$-18.8$&$3+ $&$>0.09$& & & Olsen\\
NGC 247  &$-20.5$&$26+$&$>0.16$& & & Olsen\\
NGC 300  &$-19.0$&$12+$&$>0.03$& & & Olsen\\
NGC 5170 &$-22.0$&$815\pm320$ & $1.3\pm0.5$ & & &\cite{fis90}\\
NGC 2403 &$-19.5$& 8 & & & &\cite{har91,b84}\\
NGC 2683 &$-20.4$ & $321\pm108$& $2.2\pm0.8$ & & & \cite{hhb85}\\
NGC 4569 &$-21.7$&$1000\pm400$& $2.1\pm 0.8$ & & & \cite{az98,han77}\\
NGC 4216 &$-21.8$&$700\pm380$& $ 1.3\pm0.7$ &  & & \cite{az98,han77}\\
\noalign{\smallskip}
\hline
\noalign{\smallskip}
\end{tabular}
\end{center}
\label{bigtab}
\end{table}

To answer this question, I've tried to compile the GCS \lq vital statistics'
for all of the galaxies mentioned in this review. Some of the answers
are not well-known -- for example, detailed information on the GCS metallicity
distribution is available for only a few galaxies. The total magnitude of the Milky 
Way is also a difficult quantity to measure. The values compiled,
and their sources, appear in Table~\ref{bigtab}.\footnote{Notes: Chandar et al. \cite{cft2}
used $M_B$ instead of $M_V$ to compute $S_N$ for M81; the value computed with $M_V$ is given here.
The values for NGC~7814 assume the distance modulus $m-M=31.52$.}

The specific frequency of spiral GCSs does not seem to vary as much 
as for ellipticals, although many of the values are still maddeningly uncertain.
From the available data, it seems clear that the Milky Way $S_N$ is 
not abnormally high, and might even be somewhat low. The metallicity
distributions and GCLFs are remarkably similar between the various galaxies.
M33 stands out in all of these properties, but then it is really a different
sort of galaxy from the others  -- for example, it also lacks a massive
central black hole \cite{geb01} and a central bulge.
Forbes (these proceedings; see also \cite{fbl01}) argues that 
the GCSs of spiral {\em bulges} are more similar to each other
and to ellipticals' GCSs than are spiral GCSs overall.
This would certainly explain why M33's GCS is so discrepant,
but the differences between the non-bulge GCs of spirals still
remain to be explained. The absence from the Milky Way of the young
or intermediate--age clusters so common in other galaxies is also
still a mystery.

The next time a review like this gets written, I expect that some
of the questions raised above will have been answered. Enough spirals
will have measurements of $S_N$ that its variation with galaxy/bulge 
properties can be addressed. Spectroscopic observations of GCs
in galaxies like M81 and M104 will increase the sample of galaxies
with reliable measurements of metallicity distributions and kinematics.
More detailed studies of Local Group clusters, including color--magnitude
diagrams and high--resolution spectroscopy, will show whether
the Local Group GCs are truly part of the same few--parameter
family.

\section{Summary}

The globular cluster systems of spiral galaxies are poorly-studied
compared to those of ellipticals, mostly because doing so is much more
difficult. The handful of well-studied galaxies show many similarities
among their properties, and also many similarities to elliptical GCSs.
High-resolution imaging from space has yielded much new information
about the clusters in many of these galaxies; combined with
spectroscopy from 8-m telescopes, the prospects for a more detailed
understanding of spiral galaxy GCSs seem bright.

\vspace{0.5cm}
Financial support to attend the meeting by the Smithsonian Institution
and ESO is gratefully acknowledged. I also thank R. Chandar for commenting
on a draft of the manuscript.

%


\begin{thebibliography}{8.}
\addcontentsline{toc}{section}{References}

\bibitem{az98}
K.A.~Ashman, S.E.~Zepf: {\em Globular Cluster Systems}
(Cambridge University Press, Cambridge 1998)

\bibitem{bbo98}
B.~{Barbuy}, E.~{Bica}, S.~{Ortolani}: A\&A \textbf{333}, 117 (1998)

\bibitem{hst_2}
P.~Barmby, S.~Holland, J.P.~Huchra: AJ \textbf{123}, 1937 (2002)

\bibitem{bh00}
P.~Barmby, J.P.~Huchra: ApJ \textbf{531}, L29 (2000)

\bibitem{hst_1}
P.~Barmby, J.P.~Huchra: AJ \textbf{121}, 2458 (2001)

\bibitem{bhb01}
P.~Barmby, J.P.~Huchra, J.P.~Brodie: AJ \textbf{121}, 1482 (2001)

\bibitem{b00}
P.~Barmby, J.P.~Huchra, J.P.~Brodie, D.A.~Forbes, L.L.~Schroder, C.J.~Grillmair: AJ \textbf{119}, 727 (2000)

\bibitem{b84}
P.~Battistini, F.~B\`{o}noli, L.~Federici, F.~Fusi~Pecci, R.G.~Kron: A\&A \textbf{130}, 162 (1984)

\bibitem{bs00}
M.A.~Beasley, R.M.~Sharples: MNRAS \textbf{311}, 673 (2000)

\bibitem{bks96}
F.~Bresolin, R.C.~Kennicutt, P.B.~Stetson: AJ \textbf{112}, 1009  (1996)

\bibitem{bri97}
T.J.~Bridges, K.M.~Ashman, S.E.~Zepf, D.~Carter, D.A.~Hanes, R.M.~Sharples, 
J.J.~Kavelaars: MNRAS \textbf{284}, 367 (1997)

\bibitem{cot99}
P.~C{\^ o}t{\' e}: AJ \textbf{118}, 406 (1999)

\bibitem{cbf1}
R.~Chandar, L.~Bianchi, H.C.~Ford: ApJS, \textbf{122} 431 (1999)

\bibitem{cbf2}
R.~Chandar, L.~Bianchi, H.C.~Ford: ApJ \textbf{517}, 668 (1999)

\bibitem{cbf3}
R.~Chandar, L.~Bianchi, H.C.~Ford: PASP \textbf{111}, 794 (1999)

\bibitem{cbf4}
R.~Chandar, L.~Bianchi, H.C.~Ford: A\&A \textbf{366}, 498 (2001)

\bibitem{cbf5}
R.~Chandar, L.~Bianchi, H.C.~Ford, A.~Sarajedini: ApJ \textbf{564}, 712 (2002)

\bibitem{cft1}
R.~Chandar, H.C.~Ford, Z.~Tsvetanov: AJ \textbf{122}, 1330 (2001)

\bibitem{cft2}
R.~Chandar, Z.~Tsvetanov, H.C.~Ford: AJ \textbf{122}, 1342 (2001)

\bibitem{clem01}
G.~Clementini, L.~Federici, C.~Corsi, C.~Cacciari, M.~Bellazzini, H.A.~Smith: ApJ \textbf{559}, L109 (2001)

\bibitem{xray01}
R.~{Di Stefano}, A.K.H.~Kong, M.R.~Garcia, P.~Barmby, J.~Grenier, S.S.~Murray, F.A.~Primini: ApJ \textbf{570}, 618 (2002)

\bibitem{db01}
C.M.~Dutra, E.~{Bica}: A\&A \textbf{376}, 434 (2001)

\bibitem{fis90}
P.~{Fischer}, J.E.~Hesser, H.C.~Harris, G.D.~Bothun: PASP \textbf{102}, 5 (1990)

\bibitem{fbl01}
D.A.~Forbes, J.P.~Brodie, S.S.~Larsen: ApJ \textbf{556}, L83 (2001)

\bibitem{geb01}
K.~{Gebhardt} et~al.: AJ \textbf{122}, 2469 (2001)

\bibitem{han77}
D.A.~Hanes: MemRAS \textbf{84}, 45 (1977)

\bibitem{hhb85}
H.C.~Harris, J.E.~Hesser, G.D.~Bothun, D.A.~Hanes, W.E.~Harris: AJ \textbf{90}, 2495 (1985)

\bibitem{har91}
W.E.~Harris: ARA\&A \textbf{29}, 543 (1991)

\bibitem{h96}
W.E.~Harris: AJ \textbf{112}, 1487 (1996)

\bibitem{h00}
W.E.~Harris: In: {\em Saas-Fee Advanced Course 28, 
 Stellar Evolution in Star Clusters}, ed. by B.~Binggeli, R.~Buser (Springer-Verlag, New York, 2000) pp1

\bibitem{hr00}
M.~{Hilker}, T.~{Richtler}: A\&A \textbf{362}, 895 (2000)

\bibitem{hub32}
E.P.~Hubble: ApJ \textbf{76}, 44, 1932.

\bibitem{hur00}
R.L.~Hurt, T.H.~Jarrett, J.D.~Kirkpatrick, R.M.~Cutri, S.E.~Schneider,
  M.~Skrutskie, W.~{van Driel}: AJ \textbf{120}, 1876 (2000)

\bibitem{kp99}
M.~Kissler-Patig, K.M.~Ashman, S.E.~Zepf, K.C.~Freeman: AJ \textbf{119}, 197 (1999)

\bibitem{lar02}
S.S.~Larsen, J.P.~Brodie, M.A.~Beasley, D.A.~Forbes: AJ \textbf{124}, 828 (2002)

\bibitem{m33_hires}
S.S.~Larsen, J.P.~Brodie, A.~Sarajedini, J.P.~Huchra: AJ in press (astro-ph/0208169) (2002)

\bibitem{lar01b}
S.S.~Larsen, D.A.~Forbes, J.P.~Brodie: MNRAS \textbf{327}, 1116 (2001)

\bibitem{lee01}
M.G.~Lee, S.C.~Kim, D.~Geisler, J.~Seguel, A.~Sarajedini, W.E.~Harris: In:
{\em Extragalactic Star Clusters, IAU Symposium 207}, ed. by E.K.~Grebel, D.~Geisler, 
D.~Minniti (ASP, San Francisco, 2002)

\bibitem{lmc00}
S.~{Leon}, G.~{Meylan}, F.~{Combes}: A\&A \textbf{359}, 907 (2000)

\bibitem{ll17}
K.~{Lundmark}, B.~{Lindblad}: ApJ \textbf{46}, 206 (1917)

\bibitem{ma3}
J.~{Ma}, X.~{Zhou}, J.~{Chen}, H.~{Wu}, Z.~{Jiang}, S.~{Xue}, J.~{Zhu}: AJ \textbf{123}, 3141 (2002)

\bibitem{mas}
P.~{Massey}, P.W.~{Hodge}, S.~{Holmes}, G.~{Jacoby}, N.L.~{King}, K.~{Olsen},
  A.~{Saha}, C.~{Smith}: BAAS \textbf{199}, 130.05 (2001)

\bibitem{mcd02}
J.C.~{McDowell}.
 personal communication (2002)

\bibitem{mcl00}
D.E.~{McLaughlin}: ApJ \textbf{539}, 618 (2000)

\bibitem{mey01}
G.~{Meylan}, A.~{Sarajedini}, P.~{Jablonka}, S.~G. {Djorgovski}, T.J.~{Bridges}, 
R.M.~{Rich}: AJ \textbf{112}, 830 (2001)

\bibitem{ode02}
M.~{Odenkirchen}, E.K.~{Grebel}, W.~{Dehnen}, H.W.~{Rix}, C.M.~{Rockosi},
  H.~{Newberg}, B.~{Yanny}: BAAS \textbf{200}, 10.01  (2002)

\bibitem{pbh95}
J.-M.~Perelmuter, J.P.~Brodie, J.P.~Huchra:  AJ \textbf{110}, 620 (1995)

\bibitem{pr95}
J.-M.~Perelmuter, R.~Racine: AJ \textbf{109}, 1055 (1995)

\bibitem{per02}
K.M.~{Perrett}, T.J.~{Bridges}, D.A.~{Hanes}, M.J.~{Irwin}, J.P.~{Brodie},
  D.~{Carter}, J.P.~Huchra, F.G.~Watson: AJ \textbf{123}, 2490 (2002)

\bibitem{pio02}
G.~{Piotto} et~al.: A\&A \textbf{391}, 945 (2002)

\bibitem{ros99}
A.~Rosenberg, I.~Saviane, G.~Piotto, A.~Aparicio: AJ \textbf{118}, 2306 (1999)

\bibitem{sw02}
M.~{Salaris}, A.~{Weiss}:  A\&A \textbf{388}, 492 (2002)

\bibitem{m33_cmd1}
A.~{Sarajedini}, D.~{Geisler}, P.~{Harding}, R.~{Schommer}: ApJ \textbf{508}, L37 (1998)

\bibitem{sch02}
L.L.~Schroder, J.P.~Brodie, M.~Kissler-Patig, J.P.~Huchra, A.C.~Phillips: AJ \textbf{123}, 2473 (2002)

\bibitem{wh01}
B.F.~Williams, P.W.~Hodge: ApJ \textbf{548}, 190 (2001)

\end{thebibliography}
\end{document}